\documentclass[twocolumn,10pt,amsmath,amssymb,aps,prb,nobalance]{revtex4-2}

\usepackage{graphicx}
\usepackage{bm}
\usepackage{orcidlink}
\usepackage{hyperref}
\usepackage{booktabs}
\usepackage{siunitx}
\usepackage{endnotes}

\DeclareSIUnit\angstrom{\protect \text {\r{A}}}
\newcommand{\ev}[1]{\left\langle#1\right\rangle}
\newcommand{\dd}{\mathrm{d}}

\begin{document}

\title{Magnetostriction in the $J$-$K$-$\Gamma$ model: \\ Application of the numerical linked cluster expansion}

\author{Alexander Schwenke \orcidlink{0000-0003-4083-1039}}
\email{a.schwenke@tu-braunschweig.de}
\author{Wolfram Brenig \orcidlink{0000-0002-2968-6363}\,}
\email{w.brenig@tu-braunschweig.de}
\affiliation{Institute for Theoretical Physics,
Technical University Braunschweig, D-38106 Braunschweig, Germany}

\date{\today}

\begin{abstract}
We apply the numerical linked cluster expansion (NLCE) to study thermodynamic
properties of the proximate Kitaev magnet $\alpha$-RuCl$_3$ on the honeycomb
lattice in the presence of a magnetic field. Using the extended spin-$1{/}2$
$J$-$K$-$\Gamma$ model and based on documented exchange and magnetoelastic
coupling parameters, we present results for the internal energy, the specific
heat, and the magnetization. Moreover, the linear magnetostriction coefficient
perpendicular to the plane is calculated, which is sensitive to changes of the
in-plane spin-spin correlations. We find the magnetostriction to display a
dip-like feature, in line with the temperature dependent and field-driven
suppression of magnetic order in $\alpha$-RuCl$_3$. Our results are consistent
with previous findings, establishing NLCE also as a tool to study magnetoelastic
features of quantum magnets.
\end{abstract}

\maketitle

\section{Introduction \label{sec:intro}}

Mixing spin- with lattice-degrees of freedom is relevant to a wide range of features of magnets \cite{Lines1979}, including, e.g., thermodynamic and critical properties \cite{Zhu2003}, cooperative phenomena \cite{Cross1979}, as well as phonon and ultrasound renormalization \cite{Zherlitsyn2014}. In this context, magnetostriction, i.e., the relative change of lengths $L$, with applied magnetic fields, $\Delta L /L (H)$ is a quantity of great interest in magnets with frustration, since in these the magnetic properties can depend sensitively on the geometry \cite{Tang2024, Nomura2023, Gen2022, Miyata2021, Doerr2018, Jaime2012, Hemberger2007}.

Recently, frustrated quantum magnets with Ising interactions that are bond-dependent by virtue of the exchange quantum-chemistry have become of great interest. On the honeycomb lattice and for spin-$1{/}2$ they give rise to the celebrated Kitaev model \cite{Kitaev2006}. This model harbors one of the few known quantum spin liquids (QSL) in two dimensions (2D). This QSL displays fractionalization of spins into mobile Majorana matter and $\mathbb{Z}_{2}$ gauge flux excitations (visons) \cite{Kitaev2006, Takagi2019, Motome2020, Trebst2022}. The visons are localized in the absence of external magnetic fields. The ground state is flux-free and displays gapless fermionic quasiparticles with a Dirac spectrum. All spin correlations are short ranged \cite{Baskaran2007}.

Mott-Hubbard insulators with strong spin-orbit coupling \cite{Jackeli2009}, like $\alpha$-RuCl$_{3}$ \cite{Plumb2014}, have been suggested as candidate materials, proximate to the Kitaev model. However, non-Kitaev exchange interactions lead to zigzag antiferromagnetic order below $\qty{7.1}{\kelvin}$ in this material \cite{Cao2016}. Applying an in-plane magnetic field $H$ along the Ru-Ru bond direction, $H{\parallel}b$, first induces a transition between different interplane zigzag orderings at $H{\parallel}b = H_{c1} {\sim}\qty{6.4}{\tesla}$, before suppressing order completely \cite{Sears2017,Wolter2017}, producing a quantum disordered range of $H{\parallel}b = H_{c2} {\sim}7.1{\dotsc}\qty{11}{\tesla}{\sim} H_{c3}$ which might host the sought for low-temperature QSL \cite{Baek2017, Hentrich2018,Balz2019, Schonemann2020, Balz2021}. Excitation continua observed in several probes, both, direct, i.e., Raman \cite{Sandilands2015, Nasu2016, Wulferding2020}, inelastic neutron \cite{Knolle2014, Banerjee2016, Banerjee2017, Do2017}, and resonant X-ray scattering \cite{Halasz2016}, as well as indirect, i.e., phonon spectra \cite{Metavitsiadis2020, Ye2020, Feng2021, Li_diff_2021, Feng2022, Metavitsiadis2022, Singh2023}, and ultrasound attenuation \cite{Hauspurg2024, Hauspurg2025}, have been attributed to fractionalization in $\alpha$-RuCl$_{3}$.

In view of the potential QSL which emerges from the zigzag order by increasing temperature and magnetic field, identifying the proper low-energy Hamiltonian for $\alpha$-RuCl$_{3}$ is very important. To this end, various spin models have been suggested \cite{Kim2015, Kim2016, Winter2016, Winter2017, Ran2017, Janssen2017, Wang2017, Cookmeyer2018, Wu2018, Ozel2019, Janssen2019, Andrade2020, Kim2022, Rousochatzakis2024, Moller2025}, which include additional off-diagonal and Heisenberg exchange, $\Gamma_i$ and $J_i$, respectively, beyond the pure Kitaev coupling $K$. Each of these exchange parameters can be sensitive to stress or strain and thereby impact magnetoelastic probes. In $\alpha$-RuCl$_{3}$ this relates, e.g., to spin-phonon scattering matrix elements \cite{Metavitsiadis2020, Ye2020, Feng2021, Li_diff_2021, Feng2022, Metavitsiadis2022, Singh2023, Hauspurg2024, Hauspurg2025} and to the mixing between magnetic and phononic thermal transport modes, both, longitudinal and transverse \cite{Hentrich2018, Hentrich2019, Vinkler2018, Ye2018}. Moreover, and as one prime motivation for the present study, this is relevant to interpreting the field-induced transitions in $\alpha$-RuCl$_3$ probed by the structural Gr\"uneisen parameter and magnetostriction \cite{Gass2020, Kaib2021, Kocsis2022}.

The observed magnetostriction displays pronounced and temperature dependent anomalies in the vicinity of $H_{c1,c2}$ \cite{Gass2020, Kocsis2022}. As of today, this has been described theoretically by linear spin-wave theory \cite{Gass2020} and by exact diagonalization \cite{Kaib2021}. In this situation, and as a second motivation of this work, we aim at considering the magnetostriction of $\alpha$-RuCl$_{3}$ as a testbed for a third, and complementary computational method, namely, the NLCE \cite{Rigol2006, Rigol2007}.

The NLCE is based on the linked cluster theorem and allows calculating quantities in the thermodynamic limit without the need for finite size scaling.  In contrast to other series expansions \cite{Oitmaa2006} and also earlier LCE approaches \cite{Sykes1966}, the NLCE is non-perturbative, relying on exact diagonalization (ED) to obtain the contributions from each individual cluster to the thermodynamic limit. This method has been applied previously to analyze a wide range of spin and general quantum many body systems for their ground-state \cite{Singh2012, Ixert2016}, thermodynamic \cite{ Rigol2006, Rigol2007, Khatami2011, Khatami2011-3, Khatami2012, Hayre2013, Schaefer2020}, and dynamic \cite{ Mallayya2017, Richter2020} properties.
Beyond that, it has been used to study entanglement entropies at quantum critical points \cite{ Kallin2013, Kallin2014}, quantum phase transitions \cite{ Khatami2011-3, Applegate2012, Jaubert2015, Benton2018-2, Pardini2019}, fermions in optical lattices \cite{ Khatami2011-2, Khatami2012-2, Khatami2012-3, Hart2015}, inhomogeneous \cite{ Gan2020} and disordered systems \cite{ Abdelshafy2024}, as well as others \cite{ Yang2011, Tang2013, Benton2018, Sumeet2024}.
We note, that apart from the NLCE, another successful cluster approach to spin models exists, namely the coupled-cluster-method (CCM) \cite{Bishop2000, Farnell2009, Reuther2011}. While the CCM is an exponential-Ansatz bivariational approach, which is rather different in spirit from the NLCE, both methods share the remarkable common feature of being constructed directly to satisfy the linked-cluster theorem.

\begin{figure}[tb]
\centering\includegraphics[width=1\columnwidth]{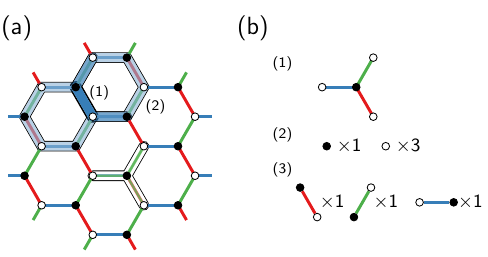}
\caption[]{(a) Honeycomb lattice and sample clusters thereon. Red, green and blue label x-, y- and z-type bond-directional exchange, respectively. Black and white vertices denote the triangular lattice basis. Smaller clusters such as (1) can be fully contained by larger ones such as (2). (b) NLCE-decomposition of a cluster into its constituent exemplified: Tricoordinated unit (1) consists of four sites (2) and, equivalently of three bonds (3), one of each type.}\label{fig:1}
\end{figure}

In this work we will use the NLCE to obtain thermodynamic properties, and in particular the magnetostriction of $\alpha$-RuCl$_{3}$. The paper is organized as follows: In Sec. \ref{sec:model} we describe the $J$-$K$-$\Gamma$ model. Sec. \ref{sec:nlce} summarizes the NLCE method and our implementation. In Sec. \ref{sec:results} we provide our results, including the energy, specific heat, and magnetization in Secs. \ref{subsec:energy}, \ref{subsec:specificheat}, and \ref{subsec:magnetization}, respectively, before discussing the magnetostriction, in \ref {subsec:magnetostriction}. We summarize in Sec. \ref{sec:summary}.

\section{\label{sec:model} Model}
We consider the spin $S{=}1{/}2$ $J$-$K$-$\Gamma$ model, which is an extension of 
Kitaev's original Hamiltonian, adapted to a more realistic description of spin-orbit
coupled magnets
\begin{equation}
  H {=} \!\!\!\sum_{
    \substack{\langle ij\rangle\parallel 
  \gamma,\\ \alpha\neq\beta\neq\gamma}} \!\!\! [J\vec{S}_i{\cdot}\vec{S}_j 
    {+} K S_{i}^{\gamma}S_{j}^{\gamma} + \Gamma (S_{i}^{\alpha}S_{j}^{\beta}
    {+} S_{i}^{\beta} S_{j}^{\alpha})]
    {+} \vec{h}{\cdot}\vec{S}\,,
\label{eq:1}
\end{equation}
where $\vec{S}_i$ are spin-$1{/}2$ operators residing on the sites $i$ of the honeycomb lattice, $\vec{S}=\sum_i \vec{S}_i$ is the total spin, and $K$ and $J$ refer to the Kitaev and Heisenberg exchanges respectively, $\Gamma$ is an off-diagonal exchange and $\vec{h}$ is an external magnetic field. $\vec{h}$ relates to the experimentally accessible field $\vec{B}$ via $\vec{h} = \mu_B g \vec{B}$ with the Bohr magneton $\mu_B$ and the $g$-tensor. The sum on $\langle i j\rangle$ extends over nearest neighbor bonds of the honeycomb lattice. The bond-directional nature of the interactions is indicated in Fig. \ref{fig:1}(a), where x-, y-, and z-type bonds are colored red, green, and blue, respectively. On a given bond along the $\gamma$-direction, the Kitaev interaction couples the $\gamma$-components of the spins, while the $\Gamma$ interaction couples the remaining two components $\alpha\neq\beta\neq\gamma$.

The $J$-$K$-$\Gamma$ model has first been suggested as a minimal model for the iridium oxide family A$_2$IrO$_3$ ($A=$ Na, Li) and can be obtained via a strong coupling expansion of a multiorbital t$_{2\textrm{g}}$ Hubbard-Kanamori Hamiltonian \cite{Rau2014}. Shortly thereafter, the same model has been proposed for $\alpha$-RuCl$_3$ \cite{Plumb2014}. Various sets of exchange parameters have been put forward in the literature \cite{Kim2015, Kim2016, Winter2016, Winter2017, Janssen2017, Ran2017, Wang2017, Cookmeyer2018, Wu2018, Ozel2019, Janssen2019, Andrade2020, Kim2022, Rousochatzakis2024, Moller2025,Kaib2021}. Here we use the set of Ref. \cite{Kaib2021} which has been obtained by extensive ab-initio analysis. We note that in the latter Ref.
further exchange interactions have also been considered. However, for the present work, and the sake of simplicity, we restrict ourselves to $J$, $K$ and $\Gamma$.
In turn, we use the following set of model parameters $\{J,K,\Gamma\} = \left\{-0.564, -1.0,0.92\right\}\times \left|K\right|$ with an overall energy scale of $\left|K\right| = \qty{10.1}{\milli\electronvolt}$.

\section{NLCE: Application to the honeycomb lattice \label{sec:nlce}}
Our method of choice, i.e., the NLCE \cite{Rigol2006} is based on the linked
cluster theorem (LCT) \cite{Oitmaa2006}, which states that for any observable
$\mathcal{O}$ representing an extensive quantity, the thermodynamic average can
be written as
\begin{equation}\label{eq:2}
\ev{\mathcal{O}}/N = \sum_{c\subset\mathcal{L}} L(c)W_{\mathcal{O}}(c)\,,
\end{equation}
where the summation extends over all connected clusters $c$
that can be embedded in the lattice $\mathcal{L}$, $N$ is the number of lattice
sites, $\langle\dots\rangle$ refers to the trace over the density matrix, and
$L(c)$ and $W_{\mathcal{O}}(c)$ are the multiplicity and the weight of a cluster
$c$, respectively.

\begin{table}[bt]
  \begin{tabular}{rr}
    \toprule
    n & Topo.\\\midrule
    1 & 1\\
    2 & 3\\
    3 & 3\\
    4 & 10\\
    5 & 18\\
  \end{tabular}
  \hspace{3em}
  \begin{tabular}{rr}
    \toprule
    n & Topo.\\\midrule
    6 & 55\\
    7 & 125\\
    8 & 360\\
    9 & 919\\
    10 & 2588\\
  \end{tabular}
  \hspace{3em}
  \begin{tabular}{rr}
    \toprule
    n & Topo.\\\midrule
    11 & 7008\\
    12 & 19756\\
    13 & 55097\\
    14 & 156357\\
    \phantom{15}\\
  \end{tabular}
  \caption[]{Number of topologically distinct clusters up to the 14th expansion 
  order on the honeycomb lattice with bond directional nearest neighbor 
  interactions}\label{tab:1}
  \end{table}

$L(c)$ is a purely geometrical quantity. It depends only on the shape of the
cluster and is given by the number of times $c$ can be embedded in
$\mathcal{L}$, without taking into account translations, i.e., per site. In
contrast to that, $W_{\mathcal{O}}(c)$ depends on the observable $\mathcal{O}$
and can be understood as the irreducible contribution of the cluster $c$ to
$\ev{\mathcal{O}}$. It is obtained recursively using the
inclusion-exclusion-principle:
\begin{equation}\label{eq:3}
W_{\mathcal{O}}(c) = \mathcal{O}(c) - \sum_{s\subset c}W_{\mathcal{O}}(s)\,.
\end{equation}
This recursion guarantees that $W_{\mathcal{O}}(c)$ comprises only correlations
such that all sites of the cluster $c$ are connected. Rearranging the LCT by 
cluster size, \eqref{eq:2} reads
\begin{equation}\label{eq:4}
\ev{\mathcal{O}}/N = \sum_{n\in\mathbf{N}}\sum_{c\in C_{n}} 
 L(c)W_{\mathcal{O}}(c),
\end{equation}
where $C_{n}$ contains all linked clusters of size $n$ embeddable in
$\mathcal{L}$ which are not related by translations. Disconnected clusters do
not contribute to \ref{eq:4} because of \ref{eq:3}.  Also note that the cluster
size $n$ is not necessarily identical to the number of sites but can rather be
defined with respect to arbitrary building blocks, e.g., bonds or plaquettes.

The NLCE then consists of truncating \ref{eq:4}, hence approximating
\begin{equation}
\ev{\mathcal{O}}/N = \sum_{n<n_{\text{max}}}\sum_{c\in C_{n}} L(c)
W_{\mathcal{O}}(c)
\label{SE}
\end{equation}
with $n_{\text{max}}$ sufficiently large and using ED to evaluate 
$W_{\mathcal{O}}(c)$ numerically.

Since clusters that host the same Hamiltonian and therefore the same density
matrix do not need to be considered individually, finding and eliminating
topologically identical clusters is crucial for lowering the computational
cost of the NLCE. In order to do so, we use the graph automorphism library
\texttt{nauty} \cite{McKay2014}.

Typically, the number of topologically distinct clusters, $n$, as well as the Hilbert
space dimension of the clusters grow exponentially with the order of the
expansion. Indeed, from Table \ref{tab:1}, for the honeycomb lattice $n\propto\mathrm{e}^{\alpha n}$ with $\alpha \approx 1.04$, for $n \gtrsim 4$. Depending on the
lattice $\mathcal{L}$ and available computational resources, this sets a limit
on the maximum order, $n_{\text{max}}$, accessible by NLCE. Yet, beyond this
truncation order, and as with all approximations that are of series expansion
type, the convergence of the bare NLCE sum of Eq. \eqref{SE} can be improved in
many cases by making use of resummation techniques \cite{Tang2013,
  Weniger1989}. In this work we employ the Euler transformation and compare
results for $n=n_{\text{max}}$ and $n=n_{\text{max}}-1$ to ensure convergence.

The Euler transform considers a sequence of finite sums
$S_k = \sum_{n=0}^{k}a_n$ and maps them to a new one, i.e.,
$\tilde{S}_j = s_{j,0}$, where $s_{j,k}$ is defined recursively by
\begin{equation}
  s_{0,k} = S_{k},\quad s_{j+1,k}=\frac{s_{j,k}+s_{j,k+1}}{2}.
\end{equation}
$\tilde{S}_j$ either converges to the same value as the original series $S_j$ 
as $j\rightarrow\infty$ or does not converge at all.

\section{\label{sec:results} Results}

In this Section we present our results for the $J$-$K$-$\Gamma$ model. This
includes the internal energy, specific heat, magnetization, and finally the
magnetostriction coefficient. One emphasis of the discussion is on analyzing
the NLCE versus different expansion orders and respective Euler transforms, in
order to study the convergence properties. Additionally we contrast the NLCE
against results from ED, obtained on finite clusters with periodic boundary
conditions. We note that in the following we refer to the magnetic field as 
$B$ to avoid confusion with the Hamiltonian $H$. However, the critical field 
values in $\alpha$-RuCl$_3$ are usually referred to as $H_{c1,2,3}$, 
therefore we will use the same notation when discussing these.

\subsection{\label{subsec:energy}Internal energy at $h=0$}

The internal energy $E=\langle H \rangle$ comprises the thermal average of the
local spin correlations accumulated by the Hamiltonian Eq. (\ref{eq:1}). The
transition into the zigzag phase in $\alpha$-RuCl$_3$ is second
order. Therefore, one expects the internal energy to be a continuous function
of $T$ even if the NLCE would converge down to $T=0$.

\begin{figure}[tb]
  \includegraphics[width=1\columnwidth]{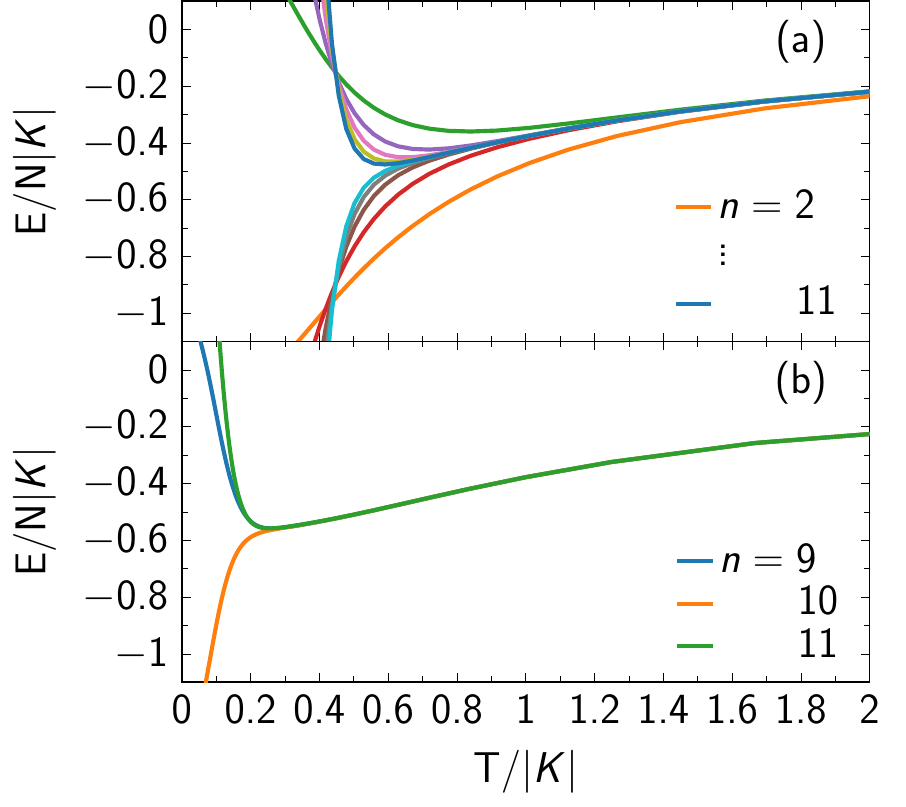}
  \caption{(a) Internal energy per site in the $J$-$K$-$\Gamma$ model 
    calculated using bare sums and (b) the Euler resummation}
    \label{fig:2}
\end{figure}

Fig. \ref{fig:2} shows $E$ per site versus temperature. As is obvious from
panel (a), up to the orders considered, the bare NLCE is converged down to
$T\approx 0.6|K|$. In order to improve this convergence of the NLCE, we apply
the Euler transform in panel (b). While the bare expansions diverge in an
alternating fashion below $T \approx 0.6|K|$, the Euler transforms of the NLCE
at orders $n = 10$ and $n = 11$ agree down to $T \approx 0.25|K|$. Speaking
differently, we may conclude that the series of Euler transformed expansions
has converged down to this temperature range. While this is a considerable
increase of the range of validity, the lowest temperature of $0.25|K|$ is
still clearly above the Ne\'el temperature $T_N\simeq \qty{7}{\kelvin}$ of
$\alpha$-RuCl$_3$. Therefore the results Fig. \ref{fig:2}(b) are well within
the paramagnetic phase.

\subsection{\label{subsec:specificheat}Specific heat at $h=0$}

We obtain the specific heat from the mean square of the energy fluctuations
\begin{equation}
  C_V = \frac{\partial E}{\partial T} = 
     \frac{\ev{H^2} - \ev{H}^2}{k_BT^2}\,.
\end{equation}
The numerator on the right-hand side is an extensive quantity and therefore
accessible to NLCE.

\begin{figure}[tb]
  \centering
  \includegraphics[width=1\columnwidth]{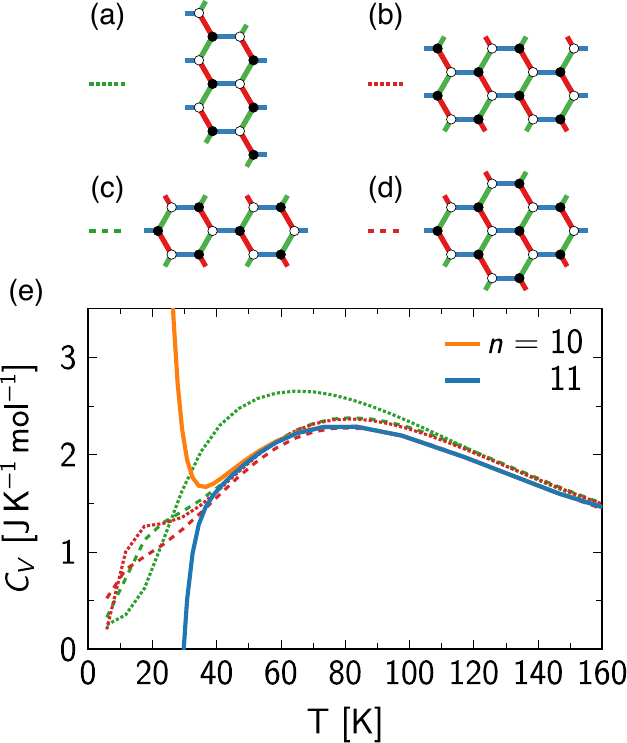}
  \caption[]{(e) Specific heat of the $J$-$K$-$\Gamma$ model: Comparison of 10th
    and 11th order NLCE (solid orange and blue), with ED on clusters of 12-sites
    (green dotted/dashed, (a)/(c)) and 16-sites (red dotted/dashed, (b)/(d)).}
  \label{fig:3}
\end{figure}

We evaluate $C_V$ up to order $n_{max}=11$ at $h=0$. In Fig. \ref{fig:3}(e) Euler resummations for $n=10,11$ are shown together with results from full spectrum ED on 12 (Fig. \ref{fig:3}(a),(b)) and 16 site (Fig. \ref{fig:3}(c),(d)) clusters using periodic boundary conditions. This figure allows to clearly observe both, strengths and weaknesses of the NLCE. Regarding the former and as determined by the convergence of the resummation, NLCE is able to reach the thermodynamic limit down to $T\gtrsim \qty{40}{\kelvin}$ on clusters of only 11 sites. This is very different for the ED. As is obvious, ED does not only display significant finite size effects, but also substantial variations versus cluster shape at fixed cluster size. Only cluster (d) is close to the thermodynamic limit within all of the range of convergence of the NLCE, requiring however a Hilbert space larger by a factor of 32. $C_V$ on this cluster properly
reproduces the position of the hump at $T\sim\qty{75}{\kelvin}$ and is only slightly too small for $\qty{40}{\kelvin}\lesssim T\lesssim \qty{70}{\kelvin}$. Clusters (b) and (c) shift the hump to a higher temperature of $T\sim\qty{85}{\kelvin}$ and overestimate $C_V$ up to $T\sim\qty{160}{\kelvin}$, while cluster (a) displays an overall shape and location of characteristic features significantly off from the NLCE and all other ED results. Cluster shapes (a) and (b) have been used in a study of the specific heat of the Kitaev-Heisenberg model, i.e., at $\Gamma=0$ \cite{Yamaji2016}. However for the parameters employed here, Fig. \ref{fig:3} demonstrates that clusters (c) and (d) are more suitable. For all remaining comparison to NLCE, we use only the larger of those clusters, i.e., (d).

As for NLCE's weakness, the low-temperature divergent behavior of the resummation below  $\qty{30}{\kelvin}$ is obvious. Here, around $T \sim \qty{10}{\kelvin} \sim 0.1|K|$ and on clusters (c)-(d) ED displays a second 'hump' of various intensity. Rephrasing all of this, NLCE typically allows reaching a controlled thermodynamic limit on clusters smaller than those required for ED. Below the low-temperature convergence however, NLCE is not defined, while ED results may still be interpreted, accepting possibly considerable finite-size effects. 

The existence of two humps in $C_V$ at $T\sim \qty{75}{\kelvin}\sim 0.65|K|$ (NLCE and ED) and at $T\sim \qty{10}{\kelvin} \sim 0.1|K|$ (ED) in Fig. \ref{fig:3} could be related to the physics of fractionalization in the pure Kitaev model \cite{Kitaev2006} at $J=\Gamma=0$. In that case the specific heat displays two maxima \cite{Nasu2014, Pidatella2019}, i.e., a high-temperature peak, which is located at $\sim\! 0.6\,|K|$ and is associated with the entropy of Majorana fermions and a low-temperature peak owing to the proliferation of gauge visons at $\sim\! 0.05\,|K|$. We caution however, that no obvious relation exists between fractionalization in the pure Kitaev model and the double peak structure in $C_V$. In fact, the so-called $\Gamma$ model for $\Gamma\neq0$ but $J=K=0$ displays similar features in $C_V$ \cite{Samarakoon2018} which persist for a broad range of $\Gamma,K\neq0$ \cite{Catuneanu2018}. Yet, the Kitaev spin liquid state is known to be severely unstable with respect to $\Gamma$ exchange interactions \cite{Rau2014}.

\subsection{\label{subsec:magnetization}Magnetization}

\begin{figure}[tb]
  \centering
  \includegraphics[width=\columnwidth]{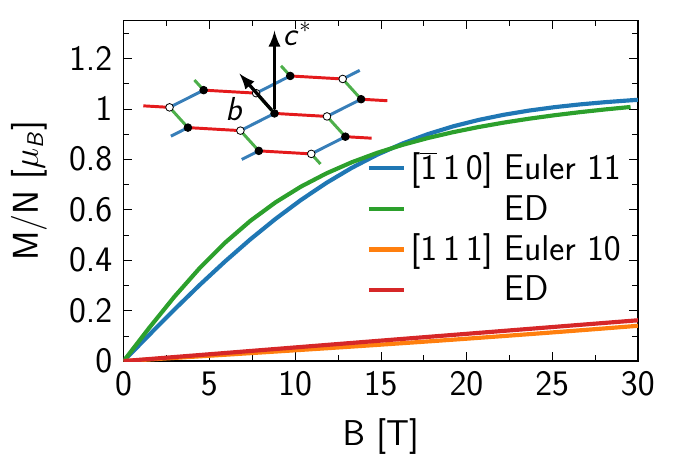}
  \caption[]{Comparing NLCE versus ED for the magnetization of the $J$-$K$-$\Gamma$
    model at $T = \qty{29.3}{K}$ for magnetic fields with in-plane, i.e.,
    $[\overline{1}\,1\,0] \hat{=} b$ and out-of-plane, i.e., $[1\,1\,1]
    \hat{=} c^{*}$ directions. ED on cluster Fig. \ref{fig:3}(d).}\label{fig:4}
\end{figure}

Experiments on $\alpha$-RuCl$_3$ have shown a significant anisotropy of the magnetic susceptibility with respect to the orientation of the applied magnetic field \cite{Johnson2015, Lampen-Kelley2018}. For a Kitaev-Heisenberg model, i.e., a $K$-$J$ model, this difference can be explained allowing for a substantial $g$-tensor anisotropy \cite{Janssen2019} which originates from strong trigonal distortion \cite{Kubota2015}. Including however an off-diagonal $\Gamma$ interaction allows for an alternative 
mechanism. In fact, $\Gamma$ acts as an antiferromagnetic coupling along the out-of-plane 
$c^*$-direction but as a ferromagnetic coupling for spins that are aligned in-plane 
\cite{Janssen2017}  - see the inset in Fig. \ref{fig:4} regarding the conventions for field 
directions. This difference automatically leads to a smaller magnetic response in the
$[1\,1\,1]$ ($c^*$) direction as compared to the $[\overline{1}\,1\,0]$ (or $b$)
direction.

Our results for the magnetization from NLCE are presented in Fig. \ref{fig:4}, where $M$ is defined as the total spin orientation along an external magnetic field $M= \langle \vec{S} \rangle \cdot \vec{B}/B$. To account for both of the previously described sources of anisotropy, we include a $g$-factor difference of $g_{c^*}=1.88$ versus $g_{b}=2.36$ \cite{Kaib2021} into \ref{eq:1}, apart from the finite off-diagonal $\Gamma$ interaction which we use for the parameters. The magnetization is calculated at a temperature of $T\sim\qty{30}{\kelvin}$. At this temperature and for $B=0$ the 10th and 11th order Euler sums are almost converged. Therefore, and since the out-of-plane magnetization remains featureless within the field range considered, we remain with NLCE at $n_{max}=10$, while for the in-plane magnetization we display $n_{max}=11$ results. For comparison, we also show magnetization from ED, which we calculate on the 16 site cluster marked (d) in Fig. \ref{fig:3}. In the out-of-plane case the agreement between NLCE and ED is excellent, whereas in the in-plane case, while still showing good agreement, the magnetization calculated by ED displays a slightly steeper slope for low magnetic field values. We mention, that increasing the magnetic field we find an improved convergence of the Euler resummation. This is due to the increasingly polarized state which can be captured on even smaller clusters. 

Obviously the magnetization versus magnetic field which we find is highly anisotropic with the in-plane magnetization being around 7-10 times larger than the out-of-plane magnetization. As a consequence the susceptibility $\partial M/\partial B$ is also anisotropic. Moreover, the in-plane magnetization starts to show saturation above approximately $\qty{15}{\tesla}$, while the out-of-plane magnetization remains linear up to $\qty{30}{\tesla}$. These results, i.e., strong anisotropy and different saturation behavior, agree very well with earlier studies using 24-site ED \cite{Winter2018,Kaib2021} on microscopic models with additional interactions compared to the ones considered here, as well as experimental results \cite{Johnson2015}.

\subsection{\label{subsec:magnetostriction}Magnetostriction}

Now that we have established that NLCE is capable of obtaining reliable results for basic thermodynamic properties of the $J$-$K$-$\Gamma$ model, we turn to the magnetoelastic property of central interest to this work, namely the linear magnetostriction coefficient $\lambda_{c^*}$. This is defined by the relative change of length along the $c^*$ direction due to the application of a magnetic field
\begin{equation}
  \lambda_{c^*} = \frac{\partial \ln{l_{c^*}}}{\partial B}\,.
\end{equation}
For the present model, magnetostriction stems from the fact that, to leading order, bond-stretching of the exchange paths leads to changes of the Hamiltonian which are proportional to spin-bilinears \cite{Zapf08}. Therefore, reciprocally, manipulating the spin-correlations by application of an external magnetic field will induce changes of the bond lengths.

Thermodynamically, the differential of the Helmholtz free energy including anisotropic
strain and stress reads
\begin{equation}
  \dd F = -S\dd T + \int\dd r^3 \sigma_{ij}\dd\epsilon_{ij} - M\dd B
\end{equation}
where $S$ is the entropy, $\sigma_{ij}$ and $\epsilon_{ij}$ are the stress and 
strain tensors respectively. Assuming homogenous stress, the spatial integral yields the volume $V$ of the crystal. Furthermore, discarding any trigonal distortion, model (\ref{eq:1}) displays $C_3^*$ symmetry, comprising combined threefold rotations in spin and real space. In turn, the stress and strain tensors are diagonal in a basis of principal axes containing $c^*$. In this basis we write
\begin{equation}
  \dd F = -S\dd T + V \sigma_{i}\dd\epsilon_{i} - M\dd B
\label{dF}
\end{equation}
with diagonal elements $\sigma_{i} \equiv \sigma_{ii}$, $\epsilon_{i} \equiv \epsilon_{ii}$ for each principal axis '$i$'. The preceding assumption of $C_3^*$ symmetry is merely used to simplify the elasticity analysis and is not meant to change the $g$-tensor anisotropy. In this work we consider uniaxial stress along $i=c^\star$. Since $\dd\epsilon_{i} = \dd\ln{l_i}$, one obtains from Eq. (\ref{dF}) the Maxwell relation
\begin{equation}
  \frac{\partial \ln{l_{c^*}}}{\partial B} = \frac{1}{V}
    \frac{\partial M}{\partial \sigma_{c^*}}.
  \end{equation}
Defining $J_\alpha = \{J,K,\Gamma\}$ for $\alpha$=1,2,3, this can be expressed via the strain dependence of $J_\alpha$, i.e., exactly the magnetoelastic coupling, as
\begin{equation}
  \lambda_{c^*} = \frac{1}{V}\sum_{\alpha}\frac{\partial M}{\partial J_\alpha}
    \frac{\partial J_\alpha}{\partial\sigma_{c^*}},
\end{equation}
where $\partial J_\alpha/\partial\sigma_{c^*}$ is of dimension $[J]/\unit{\giga\pascal}$ and to linear order is conventionally expressed using a parameter $n_\alpha$ as $n_\alpha J_\alpha$, which leads to
\begin{equation}
  \lambda_{c^*} = -\frac{1}{V}\sum_{\alpha}\left(
    \frac{\partial M}{\partial J_\alpha}n_\alpha J_\alpha\right).
  \label{eq:13}
\end{equation}
For the values of $n_\alpha$ in $\alpha$-RuCl$_3$ we rely on two inputs. I.e., we use results from {\em ab-initio} calculations under uniaxial pressure along $c^*$ \cite{Kaib2021} which can be expressed as $\{n_J, n_K, n_\Gamma\} = \{-0.283, -4.97, +1.0\}\times n_\Gamma \, \SI{}{\giga\pascal^{-1}}$ with $n_\Gamma$=0.806. For $n_\Gamma$, however, and as suggested in \cite{Gass2020}, we use a slightly larger value of 0.9 instead, in order to improve the agreement with experimental values. Finally, $V/N = \SI{92.8} {\angstrom^3}$. Note that both, positive and negative values of $n_\alpha$ occur.

\begin{figure}[tb]
  \includegraphics[width=1\columnwidth]{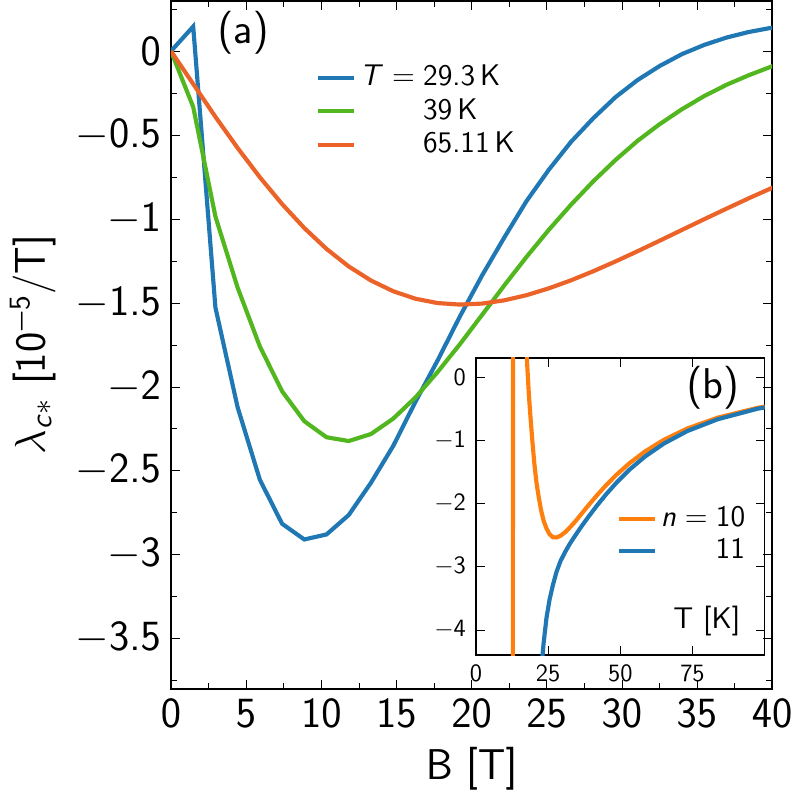}
  \caption{(a) Linear magnetostriction coefficient $\lambda_{c^\star}$ of the
    {$J$-$K$-$\Gamma$}~model from NLCE at 11th order versus in-plane magnetic field along the
    $[\overline{1}\,1\,0]$ direction for various temperatures. (b) Comparing NLCE for
    $\lambda_{c^\star}$  at 10th
    and 11th order (solid orange and blue) versus temperature
    at fixed $B=\qty{8.9}{\tesla}$.}\label{fig:5}
\end{figure}

We evaluate $\lambda_{c^*}(B,T)$ by calculating the magnetization versus temperature, magnetic field, and $J_\alpha$, using NLCE and then approximating the derivative in Eq. (\ref{eq:13}) numerically
\begin{equation}
  \frac{\partial M}{\partial J_\alpha} \approx 
  \frac{M(J_\alpha+\Delta_\alpha)-M(J_\alpha)}{\Delta_\alpha}\,,
\end{equation}
where we find a spacing of $\Delta_\alpha = J_\alpha/100$ to be sufficiently small. Fig. \ref{fig:5}(a) presents the resulting field dependence of $\lambda_{c^*}$ at $T=\qty{29.3}{\kelvin}, \qty{39}{\kelvin}$, and $\qty{65.11}{\kelvin}$. Not to clutter the figure, the plots only show Euler transforms at the largest index we obtain, i.e., $n=11$. At this order NLCE is not fully converged for all temperatures and fields used in \ref{fig:5}(a). However, the margin of error is relatively small as can be read off from Fig. \ref{fig:5}(b). This inset proves, that the uncertainty in $\lambda_{c^*}(\qty{8.9}{\tesla})$ is at most $O(10\%)$ at $T\sim\qty{29.3}{\kelvin}$ and rapidly decreases with increasing temperature. Approaching full polarization also improves convergence of the expansion, which is relevant however only at magnetic field scales beyond $\qty{20}{\tesla}$ in Fig. \ref{fig:5}(a).

Most important, and as a main finding of this work, we observe a pronounced dip of $\lambda_{c^*}$ versus magnetic field. As the temperature is lowered the dip sharpens up, deepens and its position moves to a field clearly similar to $H_{c2}$. Increasing the temperature the dip gets less pronounced and its location moves to higher fields. Qualitatively, these results regarding the field and temperature dependence of $\lambda_{c^*}$ are in agreement with experiment \cite{Gass2020,Kocsis2022} and with earlier calculations using ED \cite{Kaib2021}. Also the magnitude of $\lambda_{c^*}$ is within the experimentally observed range. We emphasize, however, that experiments have been performed for $\qty{2.4}{\kelvin}\leq T \leq \qty{50}{\kelvin}$, while 11th-order NLCE's lowest reliable temperature is $T \sim \qty{24}{\kelvin}$, which is obvious from the inset Fig. \ref{fig:5}(b). Substantially larger orders seem necessary to cover all of the experimental temperature range. {\em Cum grano salis} NLCE also agrees with conclusions from linear spin-wave calculations \cite{Gass2020}, which however, artificially, produces a divergent dip in $\lambda_{c^*}(B,T)$ at $H_{c2}$ for all temperatures.

\section{\label{sec:summary} Summary}

We have shown the NLCE to be a promising tool to study thermodynamic properties of frustrated quantum magnets including magnetoelastic phenomena. For this purpose we have analyzed the $J$-$K$-$\Gamma$ model of the Kitaev magnet $\alpha$-RuCl$_3$, contrasting results from NLCE with ED where appropriate. We find the specific heat generated by NLCE to be in the thermodynamic limit over a wide range of temperatures on clusters which are much smaller than those for ED, which, in contrast, remain subject to finite size effects for all clusters we analyzed. The magnetization we obtained from NLCE displays a strong anisotropy with respect to the direction of the applied magnetic field, which is driven by the $g$-tensor anisotropy as well as the $\Gamma$ interaction and is consistent with ED calculations and the experimental situation in $\alpha$-RuCl$_3$.

Most important, we have used NLCE to investigate the linear magnetostriction coefficient along the $c^*$-axis and found a pronounced dip at a magnetic field which we associate with the field-driven transition out of the zigzag ordered phase at $H_{c2}$ in $\alpha$-RuCl$_3$. This, as well as the temperature dependence of the magnetostriction, is in good qualitative agreement with existing theoretical analysis as well as with experimental observations.

Finally, while our results are consistent with the location of the field-driven transition in $\alpha$-RuCl$_3$, we cannot provide insight into the nature of the underlying phases. In particular regarding high fields beyond $H_{c2}$ these remain under debate, despite extensive studies \cite{Baek2017, Hentrich2018, Balz2019, Hickey2019, Schonemann2020, Balz2021}. Therefore, while it is well established that magnetic fields applied along the $b$-axis suppress the zigzag magnetic order, leading to a potentially spin-liquid-like state, the precise nature of this state and whether or not it is adiabatically connected to the high-field paramagnetic phases remain open issues.

\begin{acknowledgments}
Many stimulating discussions with Johannes Richter regarding the size-consistent cluster method for spin systems of Refs. \cite{Bishop2000, Farnell2009, Reuther2011} are gratefully acknowledged. This work has been supported in part by the DFG through Project A02 of SFB 1143 (Project-Id 247310070). Research of W.B. was supported in part by NSF grant PHY-2210452 to the Aspen Center for Physics (ACP). W.B. acknowledges kind hospitality of the PSM, Dresden.\\[1cm]
\end{acknowledgments}

\end{document}